\documentclass[12pt]{article}
\usepackage[utf8]{inputenc}
\usepackage[T1]{fontenc}
\usepackage[a4paper,margin=1in]{geometry}
\PassOptionsToPackage{hyphens}{url}
\usepackage{amsmath,amssymb,amsfonts}
\usepackage{graphicx}
\usepackage{booktabs}
\usepackage{tabularx}
\usepackage{array}
\usepackage{url}
\usepackage{xcolor}
\usepackage{setspace}
\usepackage{caption}
\usepackage{float}
\usepackage{listings}
\usepackage[numbers,sort&compress]{natbib}
\usepackage{hyperref}
\hypersetup{hidelinks}
\usepackage{tikz}
\usetikzlibrary{shapes,arrows.meta,positioning,fit,calc}

\title{CitePrism: A Transparent Hybrid AI Framework for Manuscript-Level Citation Auditing in Editorial Workflows}
\date{}

\begin{document}

\begin{center}
\vspace*{0.25em}
{\LARGE\bfseries CitePrism: A Transparent Hybrid AI Framework for Manuscript-Level Citation Auditing in Editorial Workflows\par}
\vspace{1.1\baselineskip}
{\normalsize
Gowrika Mahesh,
Budanur Madappa Darshan Gowda,
Kavana Gopladevarahalli Papegowda,
Prajwal Basavaraj,
Binh Vu,
Swati Chandna,
Mehrdad Jalali\textsuperscript{*}\par}
\vspace{0.75\baselineskip}
{\small\itshape
Applied Artificial Intelligence and Data Analytics, Department of Information, SRH University Heidelberg, Heidelberg, Germany\par}
\vspace{0.55\baselineskip}
{\small
\textsuperscript{*}Correspondence: Mehrdad Jalali, \href{mailto:mehrdad.jalali@srh.de}{mehrdad.jalali@srh.de}\par}
\end{center}
\vspace{0.9\baselineskip}

\begin{abstract}
Editors and reviewers are expected to ensure that manuscripts cite relevant, accurate, current, and ethically appropriate literature, yet manuscript-level citation auditing remains largely manual, fragmented, and difficult to scale. Citation context---not reference counts alone---determines whether a citation substantiates a claim, while incomplete or inconsistent bibliographic sources can distort automated judgments. Bibliometric and classifier-based tools advance field-level or dataset-level analysis, but rarely provide an integrated editorial workflow combining context extraction, metadata verification, self-citation review prompts, and human oversight. Large language models (LLMs) can interpret local citation neighborhoods, yet they require fact-checking, governance, and hybrid verification. We present CitePrism, a feasibility-stage hybrid decision-support \emph{prototype} for editorial citation auditing that combines LLM-assisted contextual reasoning, embedding-based semantic similarity, metadata verification, and integrity-oriented flags within a mandatory human-in-the-loop analyst workflow. CitePrism is not an autonomous misconduct detector, an automated accept/reject system, or a substitute for editorial judgment. In a \emph{single-manuscript} pilot validation ($n=104$ references; pavement engineering), agreement with human binary relevance labels reached Cohen's $\kappa = 0.429$ (moderate agreement). At operating threshold $\tau = 17$, the prototype exhibited conservative screening in that case study: all human-labeled irrelevant citations were flagged, with additional false positives requiring analyst review. These findings indicate feasibility for editorial triage assistance only; they should not be interpreted as general editorial-performance validation. Multi-manuscript, multi-annotator studies remain required.
\end{abstract}

\noindent\textbf{Keywords:} citation auditing; editorial decision support; hybrid AI; large language models; research integrity; metadata verification; human-in-the-loop systems; publication ethics; scholarly communication

\section{Introduction}

Citations are infrastructure for scholarly trust. They anchor claims in prior evidence, make intellectual lineage visible, and enable readers and reviewers to scrutinize how authors situate new contributions within existing knowledge \cite{biagioli2020gaming,anderson2021cca}. When citation practice is weak---through irrelevant or superficial citations, distorted use of prior work, omission of foundational literature, excessive self-citation, or bibliographic inaccuracy---the interpretability and credibility of a manuscript may be compromised even when its core claims appear sound.

Editorial citation checking is therefore important but under-supported. Editors and reviewers are expected to assess whether each reference is relevant to its in-text claim, whether metadata are accurate, whether self-citation is proportionate, and whether key prior work has been acknowledged \cite{cope2023authorship,seeber2020self}. This task is cognitively demanding: reference lists are long, reviewer time is limited, and citation problems are often subtle rather than obvious \cite{kovanis2016publication,anderson2021cca}.

Modern journal workflows face scale, reviewer fatigue, and integrity risks simultaneously. Submission volumes have increased across many fields \cite{bornmann2015growth,ware2015stm}, intensifying triage burdens \cite{squazzoni2017publishing}. Citation padding, coercive citation, cartels, and metadata error remain concerns \cite{wilhite2012coercive,fister2016cartels,simkin2005read,rong2026dataquality}. Editorial teams need accountable decision support that prioritizes attention without substituting for human judgment \cite{breuning2021peer,kasneci2023chatgpt}.

Existing approaches remain fragmented. Bibliometric and network analyses map co-citation structure, collaboration, and thematic concentration at field level \cite{small1973cocitation,kessler1963coupling,hassanein2025blockchain} but do not, by themselves, audit whether a specific in-text citation supports a local claim. Citation-intent and influence-classification research improves contextual and importance labeling \cite{teufel2006automatic,cohan2019structural,iqbal2026scibert}. Metadata infrastructures enable verification \cite{priem2022openalex}, yet source incompleteness and cross-database discrepancies can distort downstream analysis \cite{rong2026dataquality}. LLMs create new opportunities for contextual screening \cite{chang2024survey} but require human oversight, fact-checking, and governance safeguards \cite{kasneci2023chatgpt,ji2023survey}.

A gap remains for a transparent, editorial-facing, manuscript-level workflow integrating citation-context extraction, hybrid relevance scoring, metadata verification, self-citation review prompts, missing-literature suggestions, and mandatory human oversight. CitePrism addresses this gap as a hybrid, human-supervised decision-support prototype---not an automated misconduct detector or autonomous editorial system.

This article makes five contributions:
\begin{enumerate}
    \item A manuscript-level hybrid citation-auditing framework for editorial decision support.
    \item A transparent relevance-scoring model combining LLM-assisted contextual reasoning and embedding-based semantic similarity.
    \item A metadata-verification and integrity-flagging layer for DOI absence, metadata mismatch, missing abstracts, retraction signals, and self-citation review prompts.
    \item A human-in-the-loop editorial analyst interface supporting threshold-based triage and citation-level rationale inspection.
    \item Preliminary empirical validation on a 104-reference manuscript showing moderate agreement with human labels ($\kappa = 0.429$) and high sensitivity for irrelevant-citation detection at the selected operating point ($\tau = 17$), with conservative false-positive behavior.
\end{enumerate}

\section{Editorial Citation Auditing: Problem Context and Research Gap}

Journal editorial workflows currently lack scalable, transparent, manuscript-level tools for citation auditing. Editors and reviewers cannot feasibly inspect every reference in depth, yet citation quality materially affects peer assessment and research integrity \cite{cope2023authorship,nasem2017integrity}.

\textbf{Citation counts are insufficient.} Bibliometric indicators and network maps summarize field structure, collaboration, and concentration \cite{hassanein2025blockchain} but do not establish whether a particular citation supports the claim in which it appears \cite{anderson2021cca}.

\textbf{Citation context matters.} Anderson and Lemken argue that rigorous review requires examining how cited works are used in context---whether citations are peripheral, substantive, supportive, critical, empirical, or potentially distorted \cite{anderson2021cca}. Editorial auditing must therefore operate at claim level (in-text neighborhoods), not only at bibliography level.

\textbf{Metadata quality matters.} Citation analysis depends on complete and consistent bibliographic records. Rong et al.\ show substantial discrepancies between Web of Science and Crossref, including effects of merging sources on reference coverage, missing key nodes, and disciplinary variation in data quality \cite{rong2026dataquality}. Manuscript-level auditing must treat metadata verification as a prerequisite, not an afterthought.

\textbf{Self-citation detection is non-trivial.} Self-citation is a normal part of scholarly communication but can be misused to inflate impact \cite{seeber2020self,mondal2026credibility}. Reliable detection requires author-name disambiguation; fuzzy matching alone is vulnerable to homonyms and variant spellings \cite{mondal2026credibility}.

\textbf{LLM-only approaches are risky.} LLMs can articulate contextual fit but may hallucinate, encode bias, and produce unverifiable rationales without independent checks \cite{kasneci2023chatgpt,ji2023survey}. Kasneci et al.\ emphasize continuous human oversight, critical evaluation, and governance for responsible LLM use \cite{kasneci2023chatgpt}.

\textbf{Interpretable hybrid designs are needed.} Citation influence and relevance classification benefit from transparent rationales and human-in-the-loop verification \cite{iqbal2026scibert}. A practical editorial system should combine LLM-assisted reasoning, embedding similarity, metadata cross-checks, self-citation review prompts, and analyst-facing explainability.

CitePrism is positioned to address this gap. Its objective is editorial triage assistance: helping analysts prioritize citations for closer inspection without automating accusations, rejections, or misconduct findings.

\section{Related Work}
\label{sec:related}

Table~\ref{tab:positioning} summarizes how CitePrism relates to representative prior streams integrated in this review.

\subsection{Citation quality, citation context, and scholarly trust}
\label{sec:rw_context}

Citation quality is increasingly treated as a research-integrity issue \cite{biagioli2020gaming,wilhite2012coercive,fister2016cartels}. Beyond ethics, \emph{how} a work is cited matters. Anderson and Lemken propose citation context analysis (CCA) as a rigorous literature-review method that examines in-text citation passages to determine whether cited works are used peripherally or substantively, supported or critiqued, and whether knowledge claims have been empirically examined or distorted \cite{anderson2021cca}. CCA goes beyond raw citation counts and bibliography inspection: it requires systematic attention to local usage at claim level.

This perspective aligns with editorial citation auditing. A reference may appear in a bibliography yet fail to support the sentence in which it is invoked; conversely, a low-cited paper may be central to a specific argument. CitePrism operationalizes this insight through citation-context extraction ($\pm 1$ sentences), hybrid relevance scoring, and a rationale viewer that lets analysts inspect model explanations against context and metadata---supporting human-supervised judgment rather than automated verdicts.

\subsection{Bibliometric data quality and metadata reliability}
\label{sec:rw_metadata}

Computational citation analysis depends on high-quality bibliographic and citation data. Rong et al.\ compare Web of Science (WoS) and Crossref in a large-scale case study, showing that the sources differ in coverage of high-impact literature, reference completeness, and network structure \cite{rong2026dataquality}. Merging datasets can improve citation-network completeness in some disciplines but may also introduce low-quality links and missing key nodes, with heterogeneous effects across fields.

These findings directly motivate CitePrism's metadata-enrichment and metadata-mismatch detection layer. Parsed reference strings are cross-checked against OpenAlex and fallback APIs; discrepancies in title, year, or DOI are surfaced as analyst-review flags rather than silently overwritten \cite{priem2022openalex,rong2026dataquality}. Manuscript-level auditing must remain aware that source-data incompleteness---including missing abstracts and incomplete Crossref records---can weaken both embedding and LLM-based relevance signals.

\subsection{Citation influence, citation intent, and interpretable citation classification}
\label{sec:rw_influence}

Citation-intent corpora (e.g., SciCite, ACL-ARC) and frame-based models classify \emph{how} authors cite prior work (background, method, comparison, etc.) \cite{cohan2019structural,jurgens2018measuring,teufel2006automatic}. Recent transformer-based approaches target citation \emph{influence} or importance. Iqbal et al.\ present a SciBERT-based framework for citation influence classification with sparse rationale extraction, arguing that black-box classifiers are unsuitable for high-stakes evaluation and that human-readable rationales support verification \cite{iqbal2026scibert}.

CitePrism shares the emphasis on interpretability and human-in-the-loop review but differs in scope. Iqbal et al.\ focus on classifying citation importance within benchmark datasets; CitePrism targets manuscript-level editorial auditing, combining hybrid relevance scoring (LLM + embeddings), metadata verification, self-citation flags, missing-citation suggestions, and configurable triage thresholds for editorial workflows. Citation influence and citation \emph{relevance} to a local claim are related but not identical tasks---a limitation we discuss in Section~\ref{sec:limitations}.

\subsection{Self-citation, author-name ambiguity, and integrity screening}
\label{sec:rw_selfcite}

Self-citation is common in cumulative research programmes but can be misused to inflate metrics \cite{seeber2020self,cope2023authorship}. Ratio-only self-citation measures are insufficient because they ignore context and author identity. Mondal et al.\ propose self-citation detection combined with author-name disambiguation to assess researcher credibility on citations, noting that homonyms and spelling variants create false positives and false negatives when matching is naive \cite{mondal2026credibility}. Hybrid and ensemble strategies can improve reliability over single heuristics.

CitePrism flags \texttt{QUESTIONABLE SELF-CITE} when author overlap coincides with low relevance scores, using fuzzy name matching (\texttt{thefuzz}) as a prototype approach. These outputs are \emph{review prompts}, not misconduct determinations. Mondal et al.'s emphasis on disambiguation informs our limitation statement: robust editorial self-citation screening will require stronger identity resolution (e.g., ORCID-aware matching) in future versions.

\subsection{LLM-assisted scholarly workflows and responsible AI}
\label{sec:rw_llm}

LLMs are being applied to literature synthesis, analytics, and review support \cite{chang2024survey,liu2023llm}. Kasneci et al.\ survey opportunities and challenges of LLMs for education, highlighting benefits for engagement and personalization alongside risks of bias, brittleness, and misuse \cite{kasneci2023chatgpt}. They argue for human oversight, critical thinking, fact-checking, and clear governance---including transparency, privacy, and sustainable deployment considerations.

Although Kasneci et al.\ do not address citation auditing directly, their governance framing applies to editorial LLM use. CitePrism treats LLM outputs as hypotheses: scores and rationales require cross-checking against embeddings and external metadata, and analysts retain accountability \cite{kasneci2023chatgpt,cope2023ai,ji2023survey}. This hybrid posture contrasts with LLM-only screening pipelines.

\subsection{Bibliometric and network-based auditing perspectives}
\label{sec:rw_bibliometric}

Bibliometric mapping analyses co-citation, collaboration networks, thematic clusters, and concentration patterns to characterize research fields. Hassanein et al.\ illustrate this tradition in a bibliometric study of blockchain research in accounting and auditing, reporting collaboration structures, co-citation trends, thematic mapping, homophily, and Matthew Effect patterns \cite{hassanein2025blockchain}. Such work is valuable for field-level intelligence but does not substitute for per-citation editorial review.

CitePrism optionally includes network-style diagnostics (e.g., venue concentration) as supplementary signals, yet its core contribution is manuscript-level triage: connecting citation context, metadata integrity, and hybrid relevance to editorial decision support rather than discipline-wide mapping.

\begin{table}[H]
\centering
\caption{Positioning CitePrism relative to prior citation-analysis and editorial-support approaches.}
\label{tab:positioning}
\small
\begin{tabularx}{\linewidth}{@{}>{\raggedright\arraybackslash}p{1.55cm}>{\raggedright\arraybackslash}X>{\raggedright\arraybackslash}p{1.65cm}>{\raggedright\arraybackslash}p{1.75cm}>{\raggedright\arraybackslash}X@{}}
\toprule
\textbf{Stream / work} & \textbf{Main focus} & \textbf{Strength} & \textbf{Limitation for editorial auditing} & \textbf{CitePrism extension} \\
\midrule
CCA \cite{anderson2021cca} & In-text citation usage and rigorous reviews & Claim-level interpretive depth & Not an automated editorial workflow & Context extraction + rationale inspection \\
Data quality \cite{rong2026dataquality} & WoS vs.\ Crossref completeness & Quantifies source discrepancies & No manuscript UI or triage & Metadata verification + mismatch flags \\
Influence classification \cite{iqbal2026scibert} & SciBERT + sparse rationales & Interpretable importance labels & Benchmark classification, not editorial pipeline & Hybrid scoring + editorial triage at $\tau$ \\
Self-citation \cite{mondal2026credibility} & Disambiguation + credibility & Addresses name ambiguity & Not integrated with relevance/metadata & Fuzzy self-cite flags + relevance context \\
Responsible LLM use \cite{kasneci2023chatgpt} & Governance and oversight & Ethics framing for LLM deployment & Not citation-specific & Hybrid verification + human accountability \\
Bibliometric mapping \cite{hassanein2025blockchain} & Field-level networks/themes & Macro structure and trends & No per-claim manuscript audit & Manuscript-level screening support \\
\midrule
\textbf{CitePrism} & Editorial citation-auditing support & Integrated hybrid workflow & Preliminary single-manuscript validation & Combines rows above in one analyst-facing prototype \\
\bottomrule
\end{tabularx}
\end{table}

\section{CitePrism Framework}
\label{sec:framework}

\subsection{Design principles}
\label{sec:design_principles}

CitePrism was designed around six principles:
\begin{itemize}
    \item \textbf{Decision support, not automation.} Outputs prioritize citations for human review; they do not issue misconduct findings or accept/reject recommendations.
    \item \textbf{Transparency.} Per-citation scores, bands, rationales, and flags are inspectable; processing stages are logged.
    \item \textbf{Hybrid evidence.} LLM-assisted reasoning is combined with embedding similarity and metadata verification.
    \item \textbf{Integrity-aware signaling.} Self-citation, retraction, and metadata anomalies are surfaced as review prompts.
    \item \textbf{Configurable triage.} Analysts adjust operating threshold $\tau$ for binary flagging separately from interpretive relevance bands.
    \item \textbf{Confidentiality awareness.} Unpublished manuscripts may be processed via external APIs only under governed editorial policies; local deployment is recommended for sensitive workflows \cite{kasneci2023chatgpt,resnik2020trust}.
\end{itemize}

\subsection{System architecture}
\label{sec:architecture}

Figure~\ref{fig:editorial_positioning} situates CitePrism within a stylized editorial workflow from submission through decision, highlighting optional pre-review citation screening before formal peer review. Figure~\ref{fig:architecture} shows the modular system architecture.

\begin{figure}[H]
\centering
\begin{tikzpicture}[
    node distance=0.55cm and 0.35cm,
    box/.style={draw, rounded corners, align=center, font=\small, minimum width=2.1cm, minimum height=0.75cm},
    highlight/.style={box, fill=blue!8, thick},
    arrow/.style={-{Stealth[length=2mm]}, thick}
]
\node[box] (sub) {Manuscript\\submission};
\node[box, right=of sub] (triage) {Editorial\\triage};
\node[highlight, right=of triage] (cite) {CitePrism\\citation screening\\(optional)};
\node[box, right=of cite] (rev) {Peer\\review};
\node[box, right=of rev] (dec) {Editorial\\decision};
\draw[arrow] (sub) -- (triage);
\draw[arrow] (triage) -- (cite);
\draw[arrow] (cite) -- (rev);
\draw[arrow] (rev) -- (dec);
\node[below=0.9cm of cite, font=\small, align=center, text width=8.5cm] {Human analyst reviews flagged citations, rationales, and integrity prompts. No automated accusation or rejection.};
\end{tikzpicture}
\caption{Editorial problem context and positioning of CitePrism as optional, human-supervised citation screening before or alongside peer review.}
\label{fig:editorial_positioning}
\end{figure}

\begin{figure}[H]
\centering
\includegraphics[width=\linewidth]{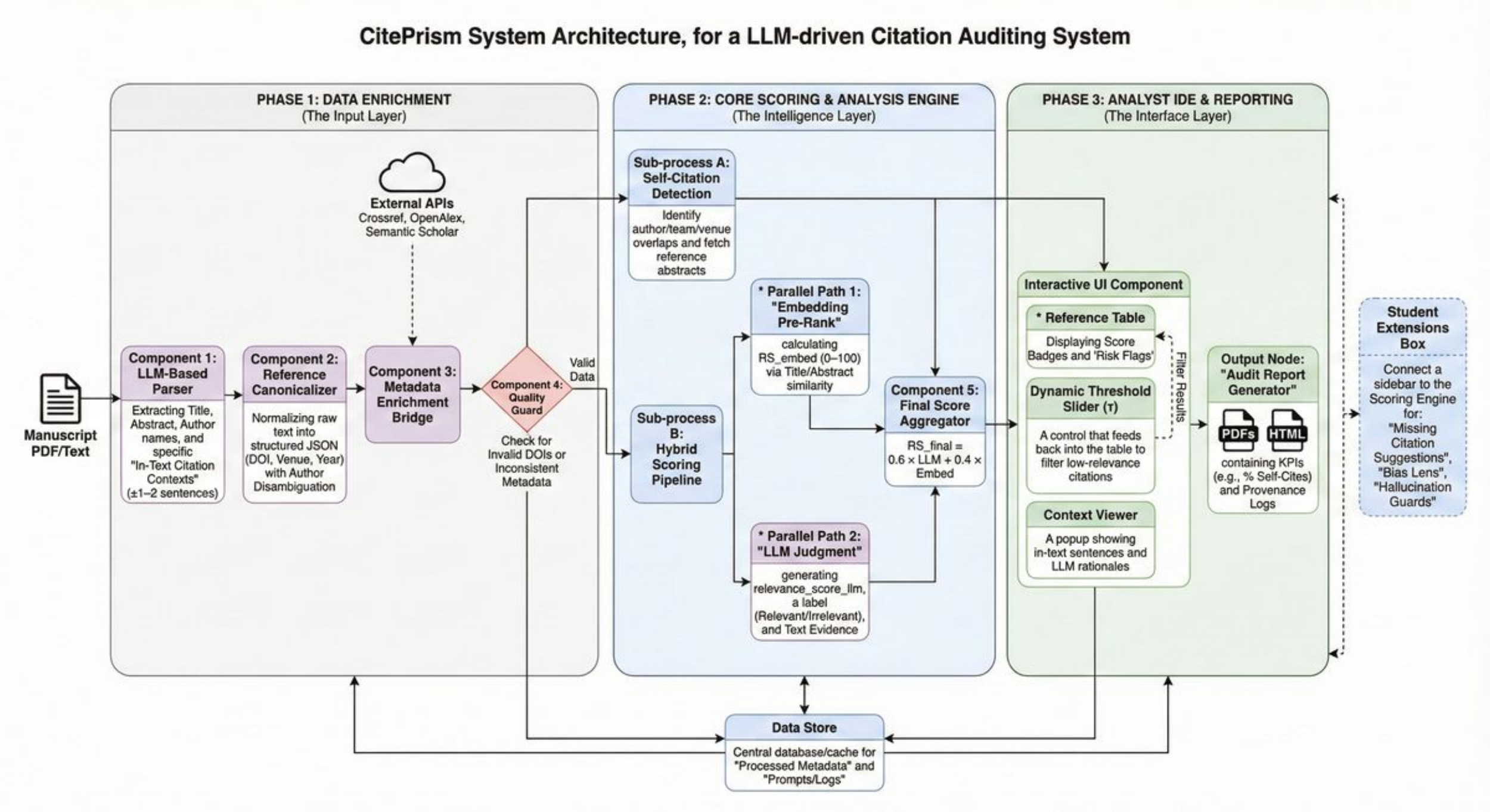}
\caption{CitePrism system architecture: five processing stages, SQLite persistence (documents, processing logs, API cache), and an editorial analyst interface. Detailed interface screenshots are provided in Appendix~A.}
\label{fig:architecture}
\end{figure}

\subsection{Manuscript parsing and citation-context extraction}
\label{sec:parsing}

\textbf{Inputs:} PDF manuscript; optional analyst configuration (reprocessing flags, threshold $\tau$).

\textbf{Outputs:} Structured manuscript metadata; reference records; in-text citation contexts (target sentence plus $\pm 1$ neighboring sentences); processing status per stage.

Following Anderson and Lemken's emphasis on citation context rather than bibliography lists alone \cite{anderson2021cca}, CitePrism extracts in-text citation neighborhoods for each reference. PDF text is extracted with \texttt{pypdf}, with \texttt{pdfminer.six} as fallback; structured parsing uses Google Gemini 2.5 Flash. For robustness on long manuscripts, parsing follows a two-mode policy: single-call processing below 60{,}000 characters and segmented processing (up to 50{,}000 characters per segment) for longer inputs. Parsed artifacts are stored in SQLite for traceability and selective reprocessing.

\subsection{Metadata enrichment and verification}
\label{sec:metadata}

In line with Rong et al.'s demonstration that WoS and Crossref differ in coverage and completeness \cite{rong2026dataquality}, CitePrism treats metadata as a first-class audit object. Parsed references are enriched primarily via OpenAlex \cite{priem2022openalex}, with Crossref among the fallbacks. When abstracts are unavailable, a four-tier strategy applies: Semantic Scholar, Crossref, arXiv, and controlled publisher-page retrieval. A metadata consistency check compares parsed and retrieved records using title similarity and year-tolerance rules. Suspected discrepancies, missing DOIs, missing abstracts, and retraction signals are retained as explicit analyst-review flags---reflecting that source-data incompleteness can distort automated relevance judgments.

\subsection{Hybrid relevance scoring}
\label{sec:scoring}

CitePrism computes two complementary signals per reference:
\begin{itemize}
    \item \textbf{Embedding score ($RS_{\mathrm{embed}}$):} Cosine similarity between manuscript and reference abstracts encoded with \texttt{all-MiniLM-L6-v2} \cite{reimers2019sentence}, rescaled to $[0,100]$.
    \item \textbf{LLM score ($RS_{\mathrm{llm}}$):} Structured batch judgments from \texttt{Llama-3.1-8B-Instruct} using manuscript abstract, citation neighborhood, and reference abstract, returning a numeric score, intent label, evidence snippet, and rationale.
\end{itemize}

The fused relevance score is:
\begin{equation}
RS_{\mathrm{final}} = 0.6 \times RS_{\mathrm{llm}} + 0.4 \times RS_{\mathrm{embed}}.
\label{eq:rsfinal}
\end{equation}

The 0.6/0.4 weighting is a \emph{prototype design choice} reflecting greater reliance on contextual LLM reasoning while retaining an independent semantic anchor; it was not optimized in the present study and should be calibrated in future multi-manuscript validation. This design parallels interpretable citation-influence classification \cite{iqbal2026scibert} but targets editorial \emph{relevance} to local claims rather than dataset-level importance labels; the two tasks should not be conflated (Section~\ref{sec:limitations}).

\textbf{Interpretive relevance bands} (on $RS_{\mathrm{final}}$): Relevant ($\geq 70$), Borderline ($40 \leq RS_{\mathrm{final}} < 70$), Irrelevant ($< 40$). These bands support qualitative interpretation.

\textbf{Operating threshold $\tau$:} A separate, analyst-adjustable cutoff for binary triage (Flagged vs.\ Clean) in the review interface and evaluation module. $\tau$ does not redefine the three bands; it controls screening sensitivity for workflow prioritization.

\begin{figure}[H]
\centering
\includegraphics[width=\linewidth]{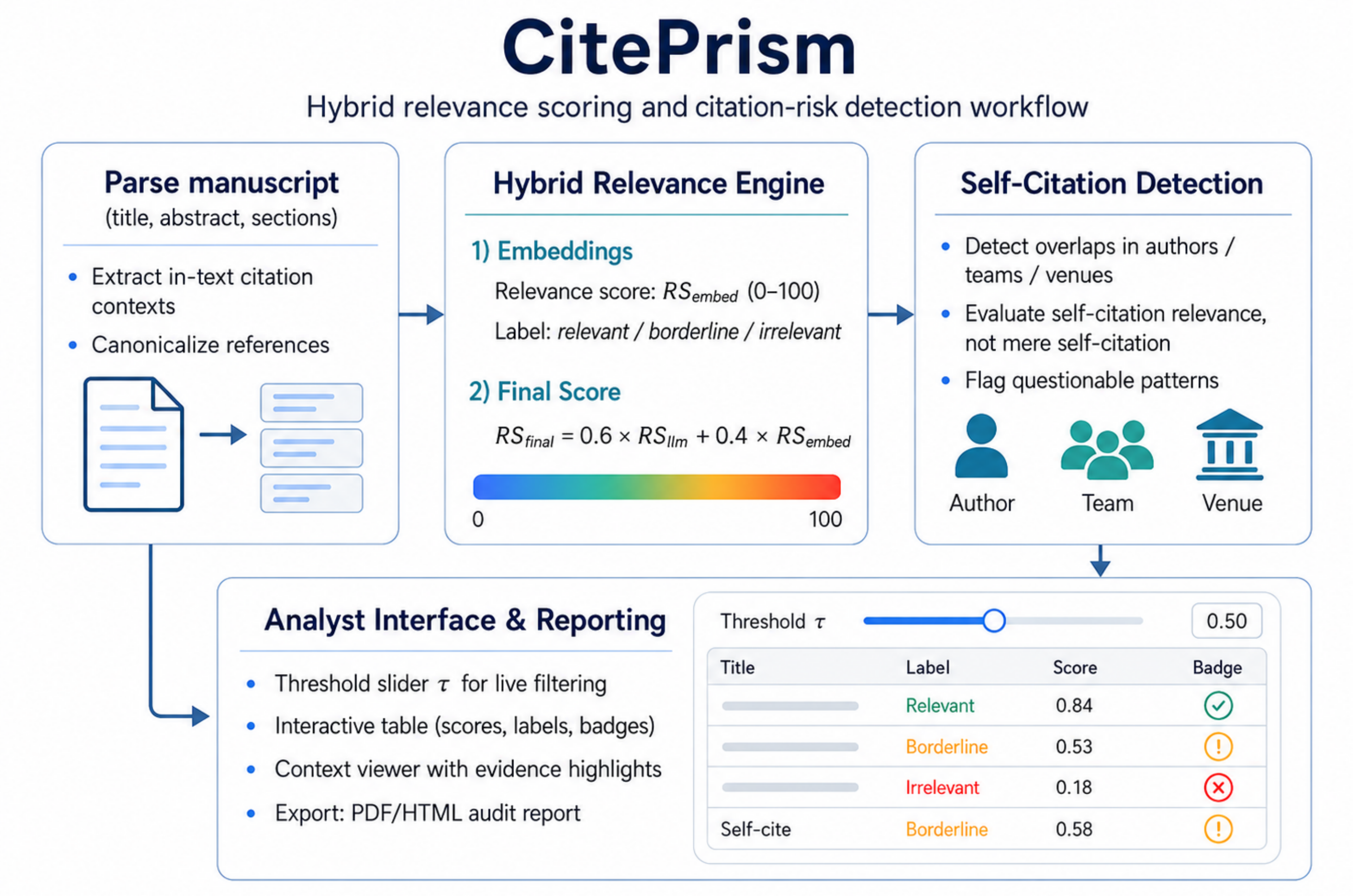}
\caption{Hybrid relevance scoring and citation-risk detection workflow from manuscript ingestion through fused scoring, band assignment, integrity flags, and analyst triage at operating threshold $\tau$.}
\label{fig:hybrid_workflow}
\end{figure}

Figure~\ref{fig:hybrid_workflow} summarizes the hybrid scoring and risk-detection workflow.

\subsection{Integrity-oriented risk flags and self-citation analysis}
\label{sec:integrity_flags}

Risk detection attaches review-oriented flags without altering $RS_{\mathrm{final}}$: \texttt{RETRACTED}, \texttt{METADATA MISMATCH}, \texttt{MISSING DOI}, \texttt{QUESTIONABLE SELF-CITE}, and missing-abstract warnings. Self-citation analysis combines fuzzy author matching (\texttt{thefuzz}) with overlap checks at author, team, and venue levels, informed by Mondal et al.'s observation that credible self-citation screening requires disambiguation beyond string overlap \cite{mondal2026credibility,seeber2020self}. Low-relevance self-citations are flagged for analyst review in line with COPE-informed interpretation \cite{cope2023authorship}; flags are review prompts, not misconduct determinations.

\subsection{Missing-citation suggestion}
\label{sec:missing}

The system proposes up to three candidate references not present in the bibliography, with short rationales, using manuscript title, abstract, and current reference list. Suggestions are generative hypotheses for expert verification and must not be treated as authoritative replacements for literature search.

\subsection{Human-in-the-loop analyst workflow}
\label{sec:hitl}

The Streamlit-based interface supports upload, staged processing, side-by-side PDF inspection, threshold adjustment, citation tables with scores and flags, rationale viewing, diagnostics, and export. Following Iqbal et al.'s argument for interpretable rationales and human verification \cite{iqbal2026scibert}, and Kasneci et al.'s insistence on oversight and critical evaluation of LLM outputs \cite{kasneci2023chatgpt}, analysts are expected to inspect flagged items, compare rationales with citation context and metadata, override model judgments, and document decisions outside the tool. Interface screenshots, runtime logs, and extended diagnostics are provided in Appendix~A (Figures~A1--A15).

\section{Experimental Setup}
\label{sec:experimental_setup}

\subsection{Case-study setting}

Validation was conducted as a \emph{pilot feasibility study} limited to one manuscript. The test manuscript (Paper~1) is a machine-learning study on resilient modulus prediction in pavement engineering with 104 references. Human annotators assigned binary labels (1 = relevant; 0 = not relevant) using manuscript context, producing \texttt{gold\_labels\_paper1.csv}. This design supports prototype feasibility assessment only; it does not establish general editorial performance, cross-domain robustness, or operational readiness for unsupervised deployment.

\subsection{Evaluation protocol}

The evaluation module aligns gold labels with scored references by reference identifier and computes Cohen's $\kappa$ \cite{cohen1960kappa}, accuracy, precision, recall, and F1 at a selected operating threshold $\tau$ \cite{landis1977kappa}. Cohen's $\kappa$ is emphasized because it adjusts for chance agreement. Three-band relevance categories remain defined on $RS_{\mathrm{final}}$ independently of $\tau$.

\subsection{Human-in-the-loop review assumptions}

Binary metrics reflect analyst-triage behavior at $\tau$, not editorial accept/reject decisions. The operating point $\tau = 17$ was selected in the prototype evaluation interface as the point reporting $\kappa = 0.429$ for Paper~1; threshold calibration across disciplines and policies remains future work.

\subsection{Governance and deployment assumptions}

The case study assumes an editorial analyst (editorial staff member, integrity officer, or designated reviewer) with authority to inspect outputs and disregard false positives. The system was not evaluated as an author-facing tool or as an automated integrity sanctioning mechanism.

\section{Results}
\label{sec:results}

\begin{figure}[H]
\centering
\includegraphics[width=0.62\linewidth]{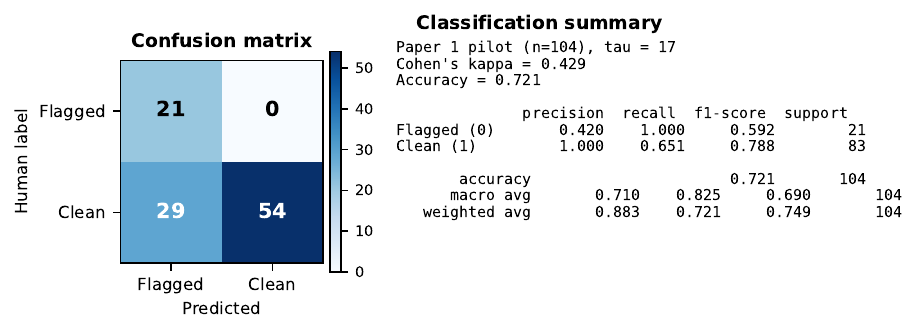}
\caption{Pilot evaluation for Paper~1 at $\tau = 17$: confusion matrix and summary metrics ($\kappa = 0.429$; $n=104$).}
\label{fig:eval_results}
\end{figure}

In this single-manuscript pilot, the prototype at $\tau = 17$ achieved Cohen's $\kappa = 0.429$, corresponding to moderate agreement on the Landis--Koch scale \cite{landis1977kappa}. Accuracy was 0.721; macro-averaged F1 was 0.690; weighted F1 was 0.749 (Table~\ref{tab:results}). Figure~\ref{fig:eval_results} shows the publication-style confusion matrix and metrics summary for this operating point. These values characterize one feasibility run and must not be read as proof of general screening accuracy across journals or domains.

\begin{table}[H]
\centering
\caption{Classification metrics for Paper~1 at operating threshold $\tau = 17$ ($n = 104$ references).}
\label{tab:results}
\begin{tabular}{lcccc}
\toprule
\textbf{Class} & \textbf{Precision} & \textbf{Recall} & \textbf{F1} & \textbf{Support} \\
\midrule
Flagged (0) & 0.420 & 1.000 & 0.592 & 21 \\
Clean (1)   & 1.000 & 0.651 & 0.788 & 83 \\
\midrule
Accuracy & \multicolumn{4}{c}{0.721} \\
Macro avg & 0.710 & 0.825 & 0.690 & 104 \\
Weighted avg & 0.883 & 0.721 & 0.749 & 104 \\
\midrule
\multicolumn{5}{l}{Cohen's $\kappa = 0.429$ (moderate agreement)} \\
\bottomrule
\end{tabular}
\end{table}

\textbf{Confusion-matrix interpretation.} All 21 human-labeled irrelevant citations were flagged (Flagged-class recall = 1.000). Twenty-nine human-labeled clean citations were also flagged (false positives). No human-labeled irrelevant citation was missed at this operating point. The system therefore exhibited \emph{conservative screening}: prioritizing sensitivity to human-identified irrelevant citations at the cost of additional analyst workload.

These results should be interpreted as \emph{pilot-stage} evidence of screening \emph{feasibility} in one domain and annotation setting, not as validation of general editorial performance, automated misconduct detection, or manuscript disposition.

\section{Discussion}
\label{sec:discussion}

\subsection{From citation counting to citation-context auditing}

Anderson and Lemken show that rigorous scholarship requires analyzing how cited works are used in context, not merely whether they appear in a reference list \cite{anderson2021cca}. Iqbal et al.\ extend this logic computationally, demonstrating that citation influence classification benefits from interpretable rationales rather than opaque scores \cite{iqbal2026scibert}. CitePrism translates these insights into an editorial workflow: extraction of local citation neighborhoods, hybrid relevance scoring, and analyst inspection of per-citation rationales. The single-manuscript pilot ($\kappa = 0.429$) suggests moderate alignment with human relevance judgments in one feasibility setting but does not resolve the inherent subjectivity of borderline citations or establish generalizable editorial performance.

\subsection{Metadata quality as a prerequisite for reliable citation auditing}

Rong et al.\ demonstrate that WoS and Crossref can diverge substantially in coverage and network completeness, and that merged datasets introduce trade-offs between breadth and precision \cite{rong2026dataquality}. In Paper~1, 43 of 104 references lacked abstracts in the test run, weakening embedding and LLM signals. CitePrism's metadata-mismatch flags are therefore not optional extras; they warn analysts when automated scores may rest on incomplete or inconsistent bibliographic ground truth.

\subsection{Self-citation flags as review prompts, not misconduct claims}

Mondal et al.\ highlight that self-citation detection for credibility assessment requires author-name disambiguation because homonyms and variants undermine naive matching \cite{mondal2026credibility}. CitePrism's \texttt{QUESTIONABLE SELF-CITE} flag combines overlap heuristics with low relevance scores to prioritize review, in line with COPE-informed practice \cite{cope2023authorship}. These signals must not be interpreted as accusations of citation manipulation; editors retain responsibility for contextual judgment, including legitimate programmatic self-citation.

\subsection{Why hybrid AI is preferable to LLM-only editorial screening}

Kasneci et al.\ caution that LLMs require continuous oversight, bias awareness, and fact-checking \cite{kasneci2023chatgpt}. Iqbal et al.\ similarly argue against black-box classification in high-stakes settings \cite{iqbal2026scibert}. CitePrism therefore combines LLM-assisted contextual scores with embedding similarity and metadata verification, exposing disagreements for analyst review. LLM-only editorial screening would risk hallucinated rationales and unverifiable triage decisions \cite{ji2023survey}.

\subsection{From bibliometric mapping to manuscript-level editorial decision support}

Bibliometric studies such as Hassanein et al.\ reveal field structure---co-citation, collaboration, thematic clusters, homophily, and Matthew Effect patterns \cite{hassanein2025blockchain}---but do not tell an editor whether a specific citation supports a claim. CitePrism complements macro mapping with micro-level triage: reference-quality checks, integrity prompts, and configurable operating thresholds for editorial workloads.

\subsection{Implications for journal editorial workflows}

In controlled \emph{pilot} use---with mandatory human oversight---CitePrism could support: (i)~\emph{editorial triage} of manuscripts with unusually large or weak bibliographies; (ii)~\emph{reviewer assistance} through structured audit reports highlighting low-relevance or metadata-anomalous citations; (iii)~\emph{reference-quality checks} before acceptance; (iv)~\emph{prioritization} of questionable self-citation patterns for human review; (v)~\emph{missing-literature suggestions} as hypotheses for expert verification; and (vi)~\emph{transparent audit logs} for internal integrity processes. None of these uses should bypass human decision-making or author communication policies.

\subsection{Interpretation of conservative screening behavior}

At $\tau = 17$, CitePrism favored false positives over false negatives: all 21 human-labeled irrelevant citations were flagged, with 29 false positives among clean labels (accuracy 0.721; weighted F1 0.749). For editorial triage, missing a problematic citation may be costlier than reviewing additional borderline cases. Threshold selection remains policy-dependent and field-specific.

\subsection{Future work}

Future validation should include: multi-manuscript and multi-domain studies; multiple independent annotators and inter-annotator agreement; stronger author-name disambiguation following Mondal et al.\ \cite{mondal2026credibility}; expanded metadata-source integration informed by Rong et al.\ \cite{rong2026dataquality}; field-specific calibration of $\tau$ and fusion weights; comparison against SciBERT-style citation-influence baselines \cite{iqbal2026scibert}; and local or private LLM deployment for confidential manuscripts \cite{kasneci2023chatgpt}.

\begin{figure}[H]
\centering
\begin{tikzpicture}[
    node distance=0.5cm and 0.4cm,
    box/.style={draw, rounded corners, align=center, font=\small, minimum width=2.0cm, minimum height=0.7cm},
    arrow/.style={-{Stealth[length=2mm]}, thick}
]
\node[box] (input) {Manuscript\\+ policy};
\node[box, right=of input] (auto) {CitePrism\\signals};
\node[box, right=of auto] (human) {Editorial\\analyst review};
\node[box, right=of human] (policy) {Journal policy and communication};
\node[box, below=0.85cm of human, fill=orange!10] (account) {Human accountability\\(no auto-sanctions)};
\draw[arrow] (input) -- (auto);
\draw[arrow] (auto) -- (human);
\draw[arrow] (human) -- (policy);
\draw[arrow] (human) -- (account);
\end{tikzpicture}
\caption{Human-in-the-loop governance model for editorial deployment: automated signals inform analyst review; editors retain accountability and policy decisions.}
\label{fig:governance}
\end{figure}
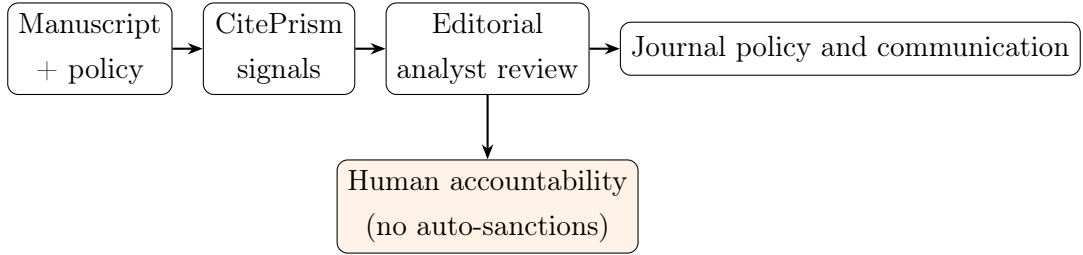

\section{Limitations}
\label{sec:limitations}

\textbf{Pilot scope.} CitePrism is a feasibility prototype, not a validated production editorial system. Results derive from one 104-reference manuscript in pavement engineering; they must not be interpreted as general editorial-performance validation.

\textbf{Single manuscript and domain.} Cross-disciplinary generalization is untested.

\textbf{Limited annotation setting.} A single gold-label file was used; inter-annotator agreement and adjudication protocols were not evaluated.

\textbf{Interpretive citation context.} Following Anderson and Lemken \cite{anderson2021cca}, relevance judgments are context-sensitive and may require domain expertise beyond automated scores.

\textbf{Bibliographic source incompleteness.} As Rong et al.\ show, Crossref and WoS differ in coverage and completeness \cite{rong2026dataquality}; missing abstracts (43/104 in Paper~1) and metadata errors can distort hybrid scoring.

\textbf{Self-citation and identity.} CitePrism lacks robust author-name disambiguation; Mondal et al.\ demonstrate that credible self-citation analysis requires stronger identity resolution than fuzzy matching \cite{mondal2026credibility}.

\textbf{Task distinction.} Citation influence/importance classification \cite{iqbal2026scibert} is related to, but not identical with, editorial relevance to a local claim; metrics from one task may not transfer to the other.

\textbf{No SciBERT baseline comparison.} The current validation does not compare CitePrism against SciBERT-style citation-classification baselines \cite{iqbal2026scibert}.

\textbf{Prototype weighting and thresholding.} The 0.6/0.4 fusion weights and $\tau = 17$ were not systematically optimized.

\textbf{LLM fact-checking requirement.} LLM rationales require analyst verification and must not be treated as authoritative \cite{kasneci2023chatgpt,ji2023survey}.

\textbf{External APIs.} Parsing and scoring rely on third-party services, raising privacy, reproducibility, and governance concerns \cite{kasneci2023chatgpt,resnik2020trust}.

\textbf{Generative suggestions.} Missing-citation proposals require independent verification.

\textbf{Multi-manuscript validation needed.} Larger studies with multiple annotators and manuscripts remain essential.

\section{Ethical, Governance, and Deployment Considerations}
\label{sec:ethics}

Human oversight is mandatory \cite{kasneci2023chatgpt,cope2023authorship}. CitePrism must not automatically accuse authors of misconduct. Integrity flags---including self-citation prompts informed by Mondal et al.'s caution about identity ambiguity \cite{mondal2026credibility}---and low relevance scores are \emph{attention-prioritization signals} only. Editors remain accountable; authors should not be penalized on automated scores alone \cite{nasem2017integrity}.

LLM-generated explanations must be checked against citation context, metadata records, and external sources \cite{kasneci2023chatgpt,iqbal2026scibert,rong2026dataquality}. Kasneci et al.\ emphasize bias awareness, fact-checking, transparency, privacy, and sustainable deployment; these principles apply directly to editorial use of LLMs on unpublished manuscripts \cite{kasneci2023chatgpt}.

Editorial offices need clear policies before pilot use: when screening occurs; which models and APIs are used; data minimization and retention; analyst training; and author communication pathways. Unpublished manuscripts require confidentiality safeguards; local or private deployment should be considered when third-party API processing is unacceptable \cite{resnik2020trust}.

CitePrism is intended for pilot-stage citation-auditing support in controlled editorial environments, not for unsupervised deployment or autonomous rejection workflows.

\section{Conclusion}
\label{sec:conclusion}

Citation auditing is a core editorial-quality and research-integrity problem, yet manuscript-level practice remains difficult to scale with manual review alone. CitePrism contributes a transparent hybrid \emph{prototype} for editorial decision support, combining LLM-assisted contextual reasoning, embedding similarity, metadata verification, integrity flags, and mandatory human-in-the-loop triage. It is not an autonomous misconduct detector or automated accept/reject system. A single-manuscript pilot ($n=104$) showed moderate agreement with human labels ($\kappa = 0.429$) and conservative screening at $\tau = 17$, capturing all human-labeled irrelevant citations in that run while generating additional false positives for analyst review. These findings support feasibility testing only and require multi-manuscript validation before stronger deployment claims.

Future work requires multi-manuscript evaluation, multiple annotators, inter-annotator agreement analysis, discipline-specific threshold calibration, ORCID-aware author disambiguation \cite{mondal2026credibility}, comparison with SciBERT-style baselines \cite{iqbal2026scibert}, richer metadata governance informed by Rong et al.\ \cite{rong2026dataquality}, and privacy-preserving deployment options \cite{kasneci2023chatgpt}. Until such evidence is available, CitePrism should be understood as human-supervised decision support that augments---rather than replaces---editorial judgment.

\section*{Acknowledgements}

The authors thank SRH University Heidelberg for academic support.

\section*{Author Contributions (CRediT)}

\textbf{Gowrika Mahesh, Budanur Madappa Darshan Gowda, Kavana Gopladevarahalli Papegowda, and Prajwal Basavaraj:} Software, data curation, visualization, validation, and writing---original draft.

\textbf{Binh Vu:} Methodological guidance, technical review, validation support, and writing---review and editing.

\textbf{Swati Chandna:} Supervision, methodological guidance, and writing---review and editing.

\textbf{Mehrdad Jalali:} Conceptualization, supervision, methodology, research-integrity framing, writing---review and editing, project administration, and corresponding-author oversight.

\section*{Conflict of Interest}

The authors declare no competing financial or non-financial interests relevant to this work.

\section*{Data and Code Availability}

The source code for CitePrism is publicly available at:\\
\url{https://github.com/SRH-Heidelberg-University-ADSA/CitePrism}

The repository includes the application source code, processing pipeline modules, Streamlit interface, SQLite schema, evaluation scripts, gold-label templates, README documentation, threshold guidance (Appendix~C), hybrid-scoring pseudo-code (Appendix~D), and a sample audit-report schema (Appendix~E). Anonymized or synthetic demonstration materials are included where legally and ethically permissible. The copyrighted case-study manuscript (Paper~1) is \emph{not} included. Interface screenshots and extended implementation evidence are provided in Appendix~A of this document.

\section*{AI and Tool-Use Statement}

Large language model (LLM) systems were used only as components of the evaluated CitePrism prototype (document parsing and structured relevance reasoning). The manuscript text, analysis, interpretation, and final approval remain the responsibility of the human authors. All LLM-generated outputs discussed in this paper were subject to human review and are treated as decision-support signals rather than authoritative editorial judgments.

\bibliographystyle{plainnat}
\bibliography{references}

\appendix
\renewcommand{\thefigure}{A\arabic{figure}}
\renewcommand{\thetable}{A\arabic{table}}
\setcounter{figure}{0}
\setcounter{table}{0}

\section{Interface and Implementation Evidence}
\label{app:interface}

This appendix provides interface screenshots and runtime diagnostics referenced in the main text. Main-text Figures~1--5 focus on editorial positioning, architecture, hybrid workflow, evaluation results, and governance.

\begin{figure}[H]
\centering
\includegraphics[width=\linewidth]{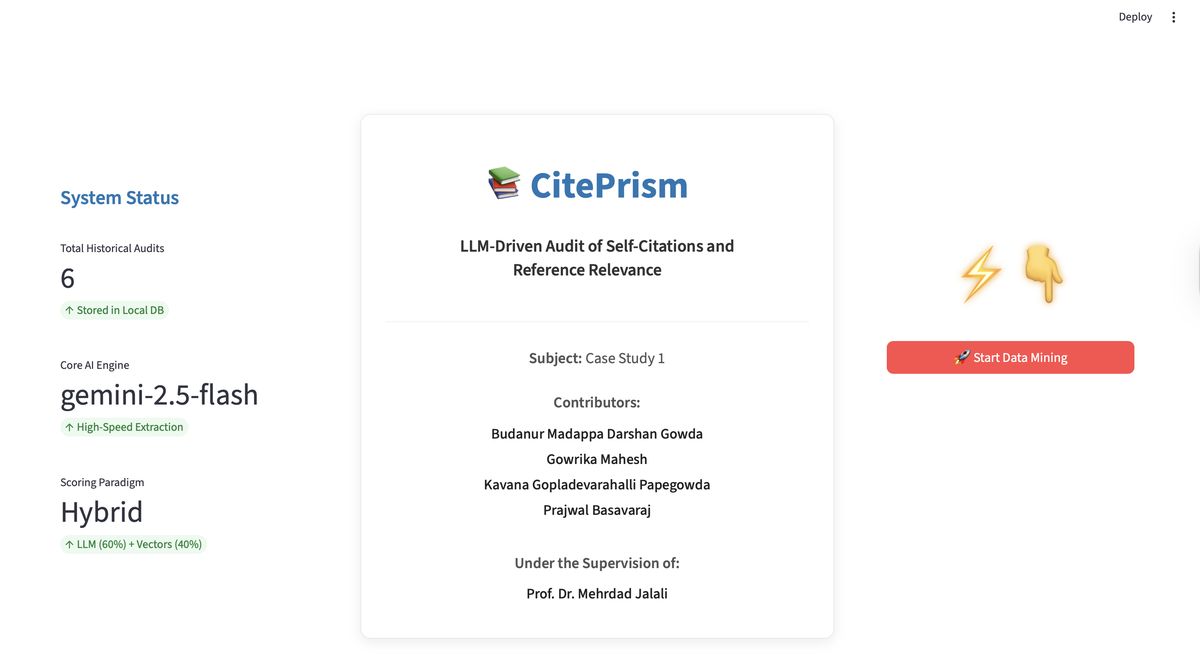}
\caption{Application landing view with system status and active hybrid-analysis configuration.}
\label{fig:a1}
\end{figure}

\begin{figure}[H]
\centering
\includegraphics[width=\linewidth]{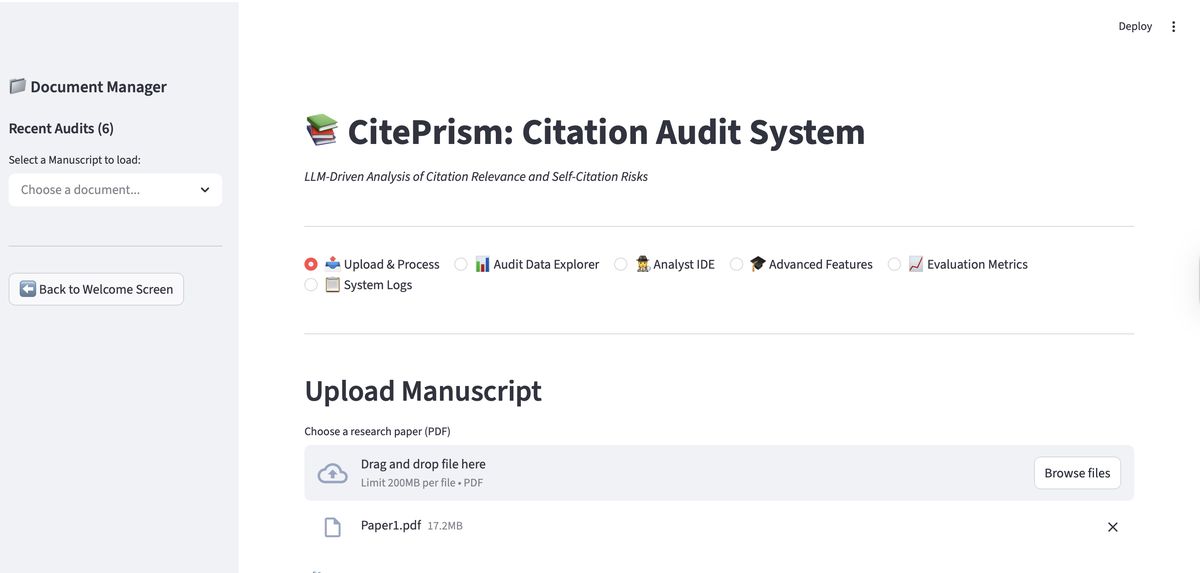}
\caption{Upload-and-process view with stage-level controls for standard or selective reprocessing.}
\label{fig:a2}
\end{figure}

\begin{figure}[H]
\centering
\includegraphics[width=\linewidth]{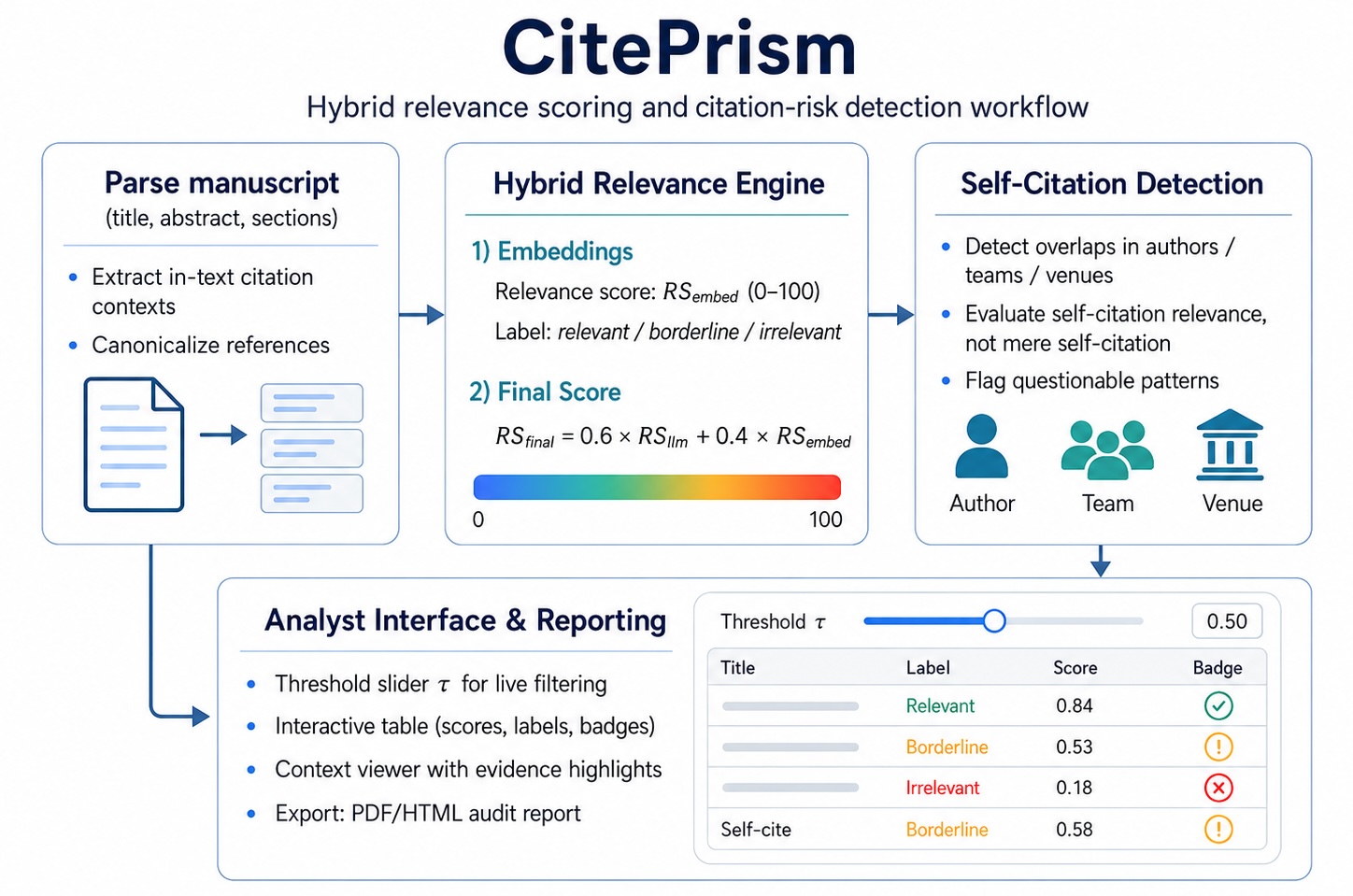}
\caption{Live extraction and scoring logs for progress monitoring and troubleshooting.}
\label{fig:a3}
\end{figure}

\begin{figure}[H]
\centering
\includegraphics[width=\linewidth]{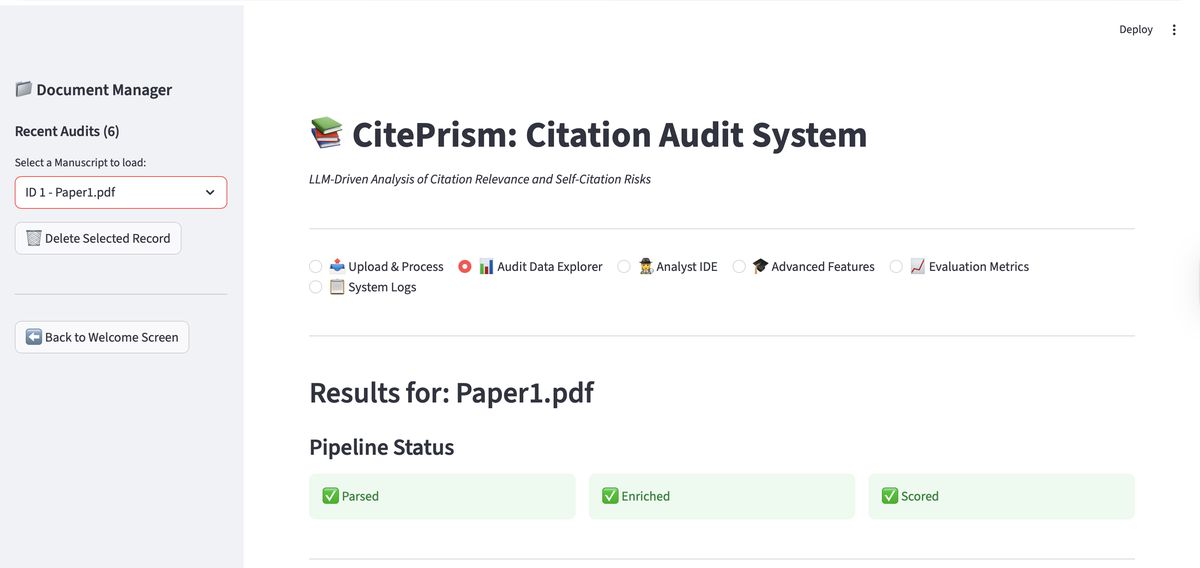}
\caption{Pipeline status panel indicating completion states for parsing, enrichment, and scoring.}
\label{fig:a4}
\end{figure}

\begin{figure}[H]
\centering
\includegraphics[width=\linewidth]{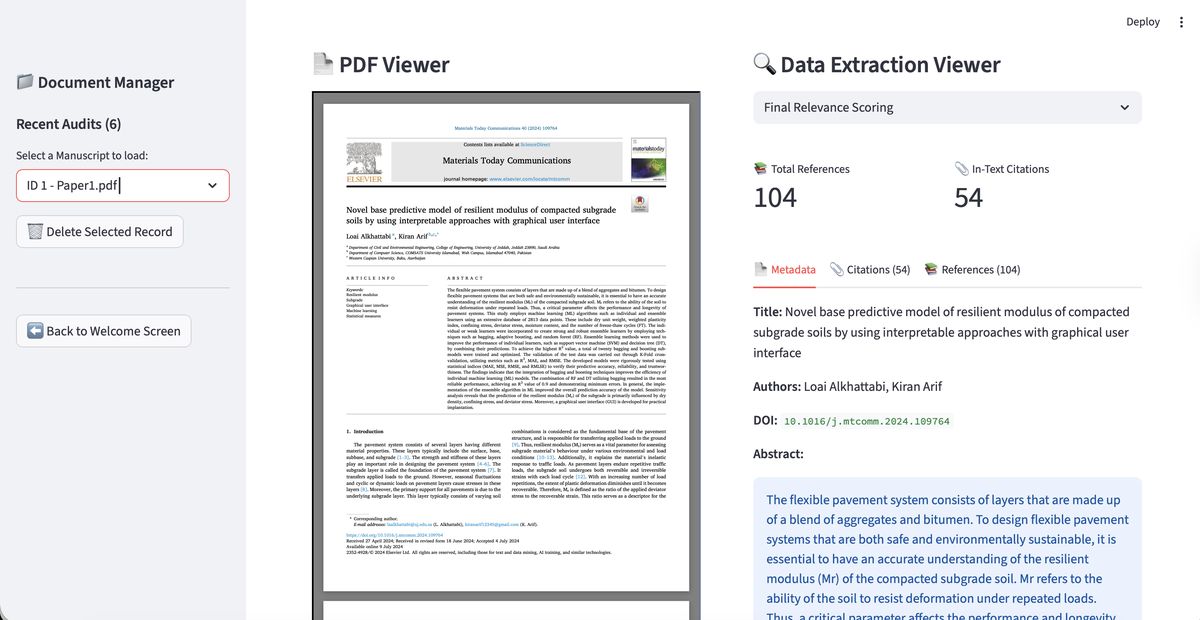}
\caption{Editorial analyst workspace with side-by-side PDF rendering and structured extraction review.}
\label{fig:a5}
\end{figure}

\begin{figure}[H]
\centering
\includegraphics[width=\linewidth]{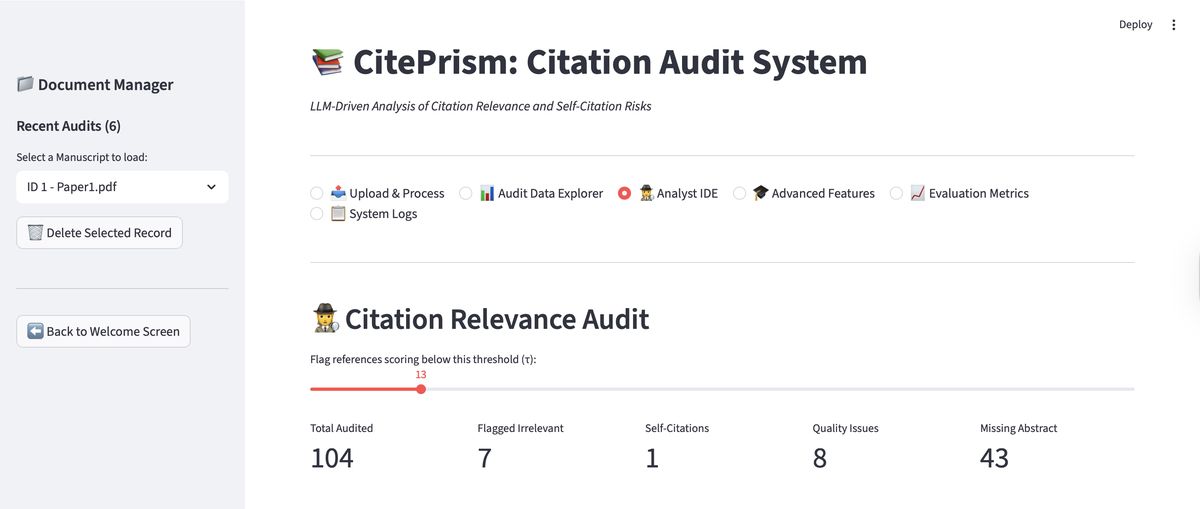}
\caption{Audit overview with headline indicators and adjustable operating threshold $\tau$.}
\label{fig:a6}
\end{figure}

\begin{figure}[H]
\centering
\includegraphics[width=\linewidth]{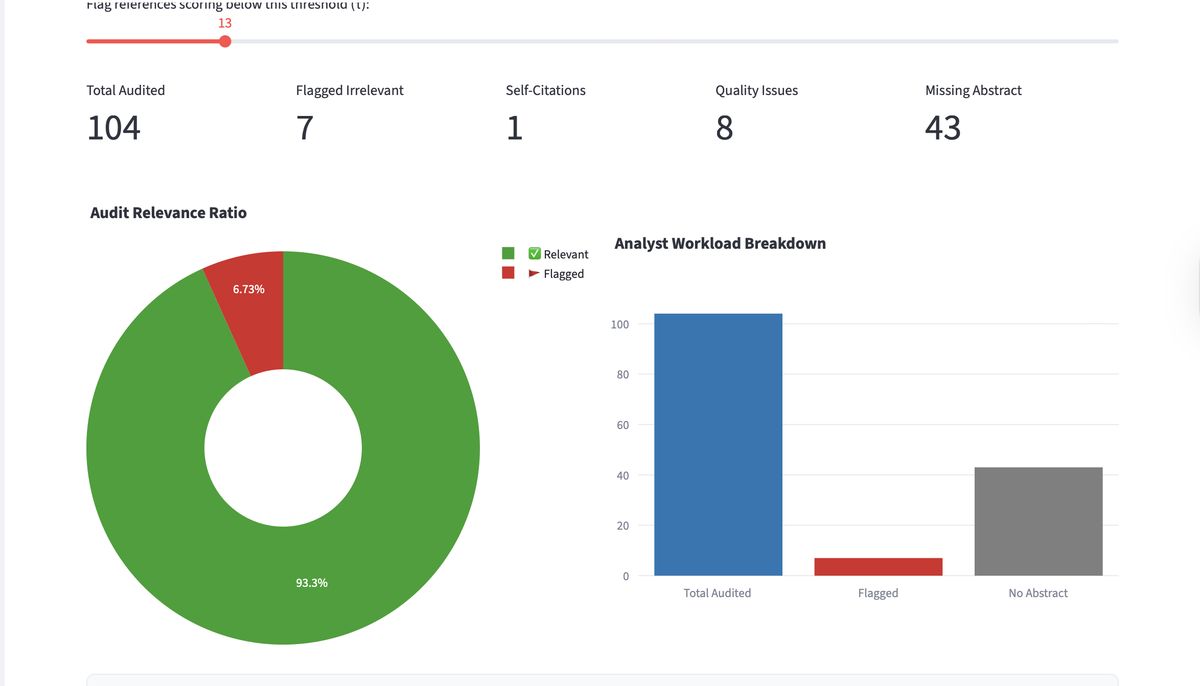}
\caption{Summary panels: relevance distribution and analyst workload profile.}
\label{fig:a7}
\end{figure}

\begin{figure}[H]
\centering
\includegraphics[width=\linewidth]{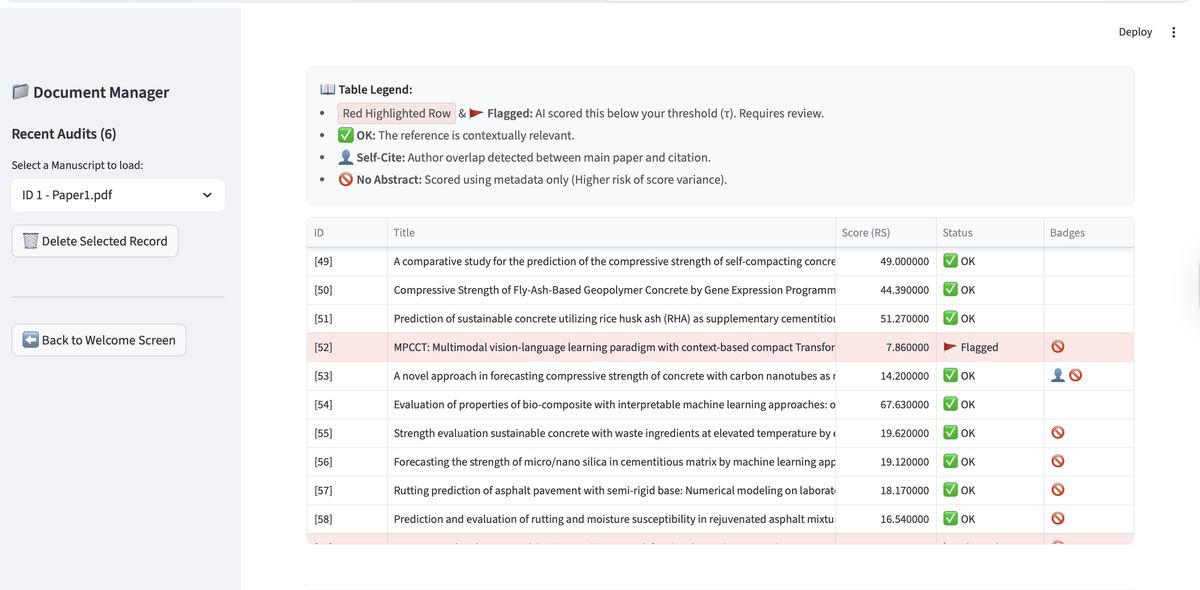}
\caption{Citation-level audit table with scores, binary flag status, self-citation markers, and missing-abstract indicators.}
\label{fig:a8}
\end{figure}

\begin{figure}[H]
\centering
\includegraphics[width=\linewidth]{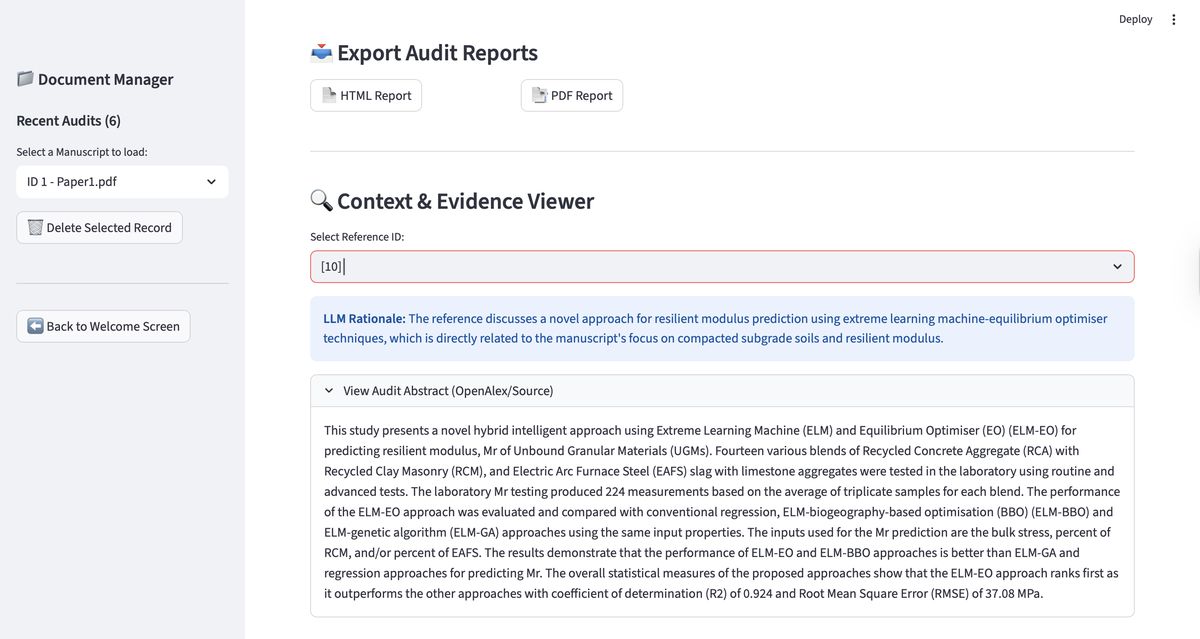}
\caption{Citation context and evidence viewer with rationale text and enriched metadata.}
\label{fig:a9}
\end{figure}

\begin{figure}[H]
\centering
\includegraphics[width=\linewidth]{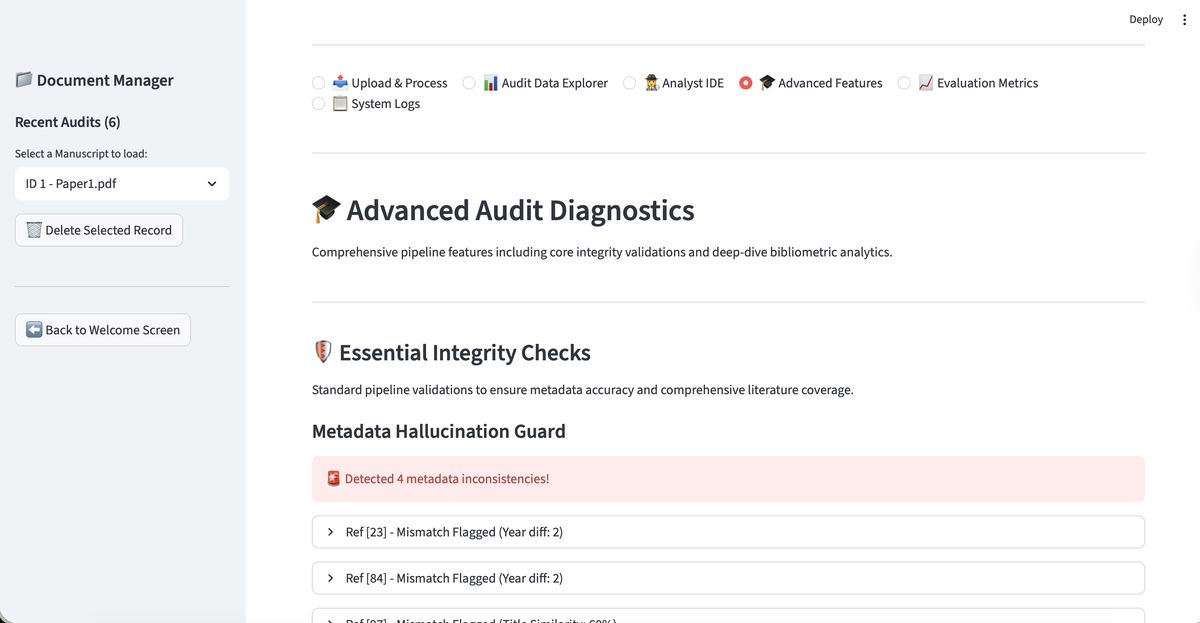}
\caption{Metadata consistency diagnostics highlighting parser-versus-metadata discrepancies.}
\label{fig:a10}
\end{figure}

\begin{figure}[H]
\centering
\includegraphics[width=\linewidth]{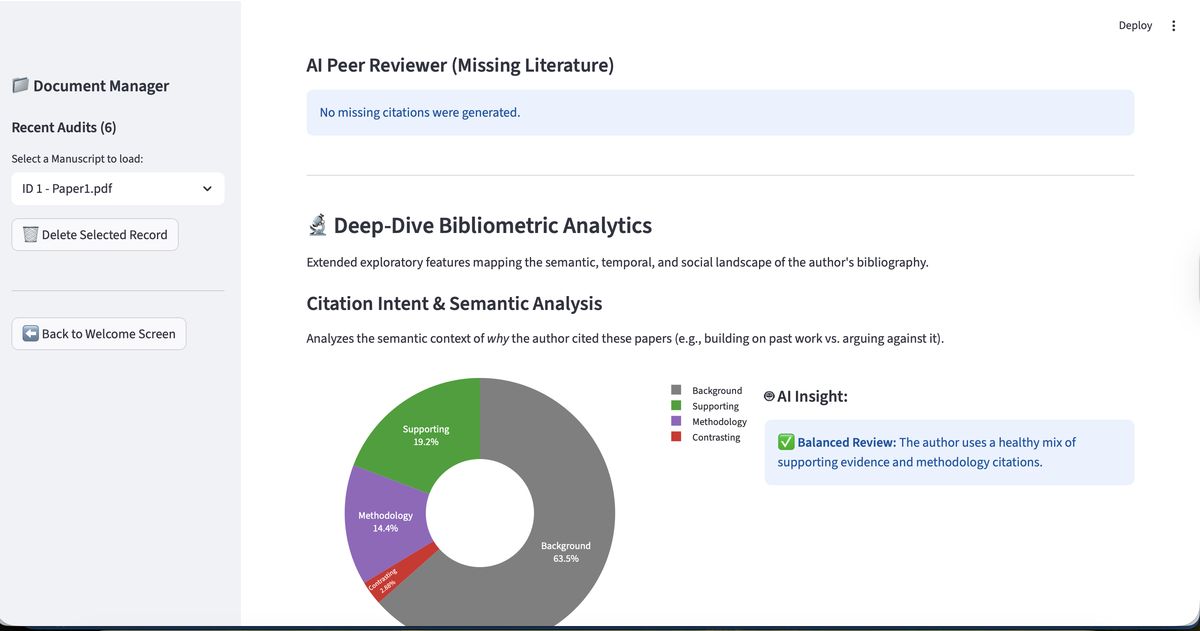}
\caption{Citation-intent distribution derived from LLM outputs.}
\label{fig:a11}
\end{figure}

\begin{figure}[H]
\centering
\includegraphics[width=\linewidth]{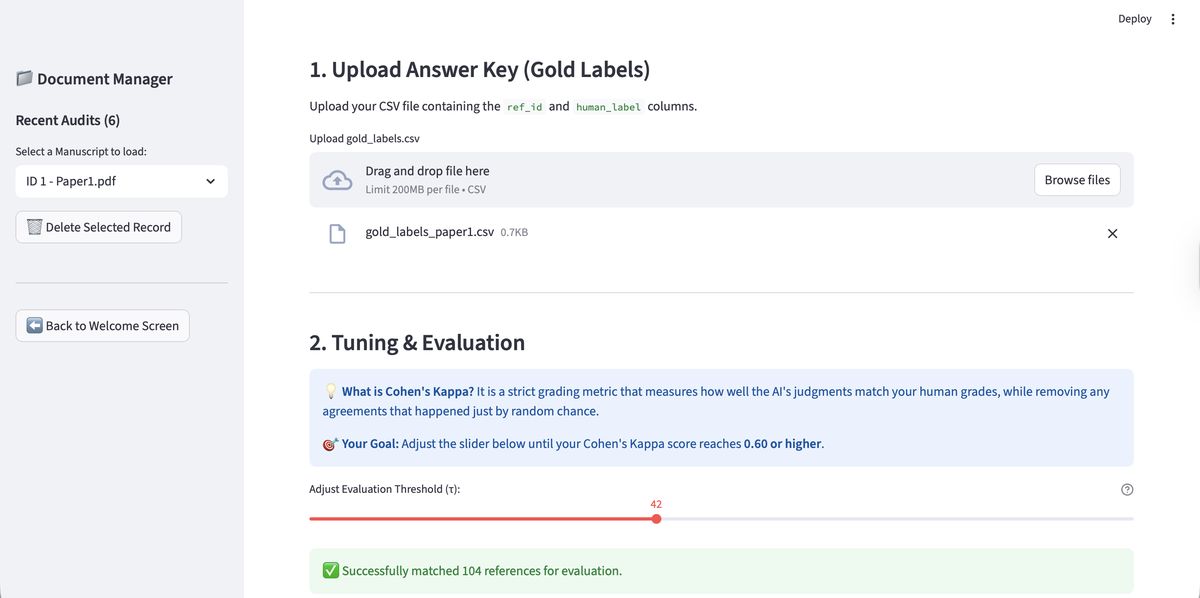}
\caption{Evaluation interface with gold-label upload and threshold control; all 104 references matched in the case study.}
\label{fig:a12}
\end{figure}

\begin{figure}[H]
\centering
\includegraphics[width=\linewidth]{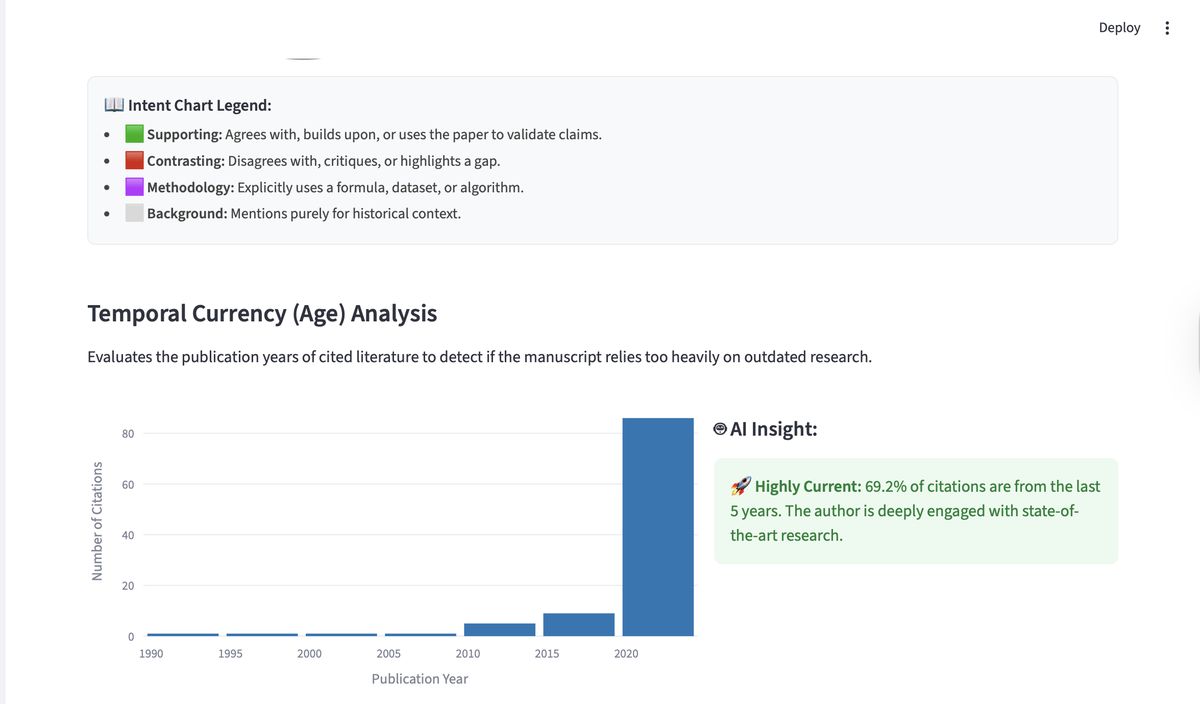}
\caption{Temporal currency analysis (69.2\% of references from the most recent five years in Paper~1).}
\label{fig:a13}
\end{figure}

\begin{figure}[H]
\centering
\includegraphics[width=\linewidth]{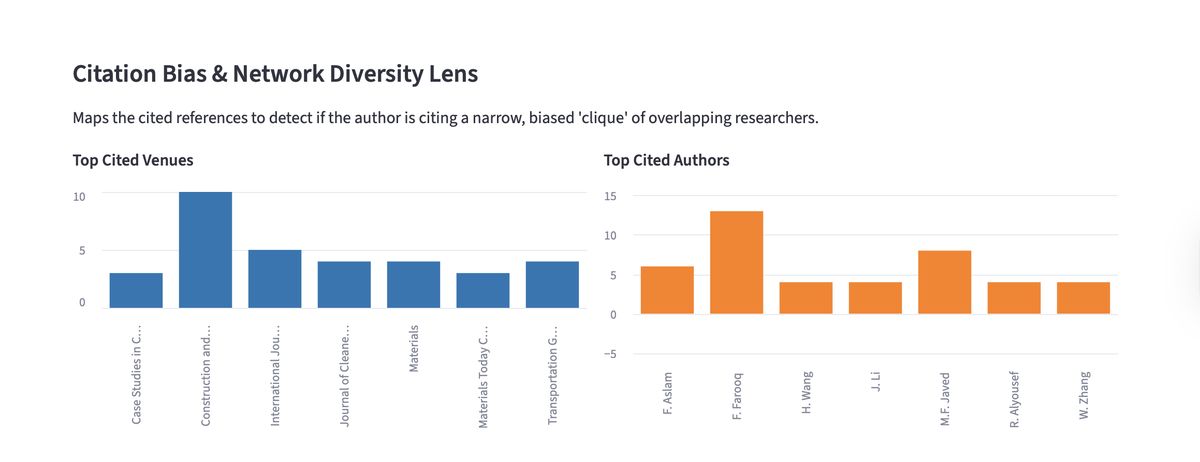}
\caption{Venue and author concentration summaries for diversity inspection.}
\label{fig:a14}
\end{figure}

\begin{figure}[H]
\centering
\includegraphics[width=\linewidth]{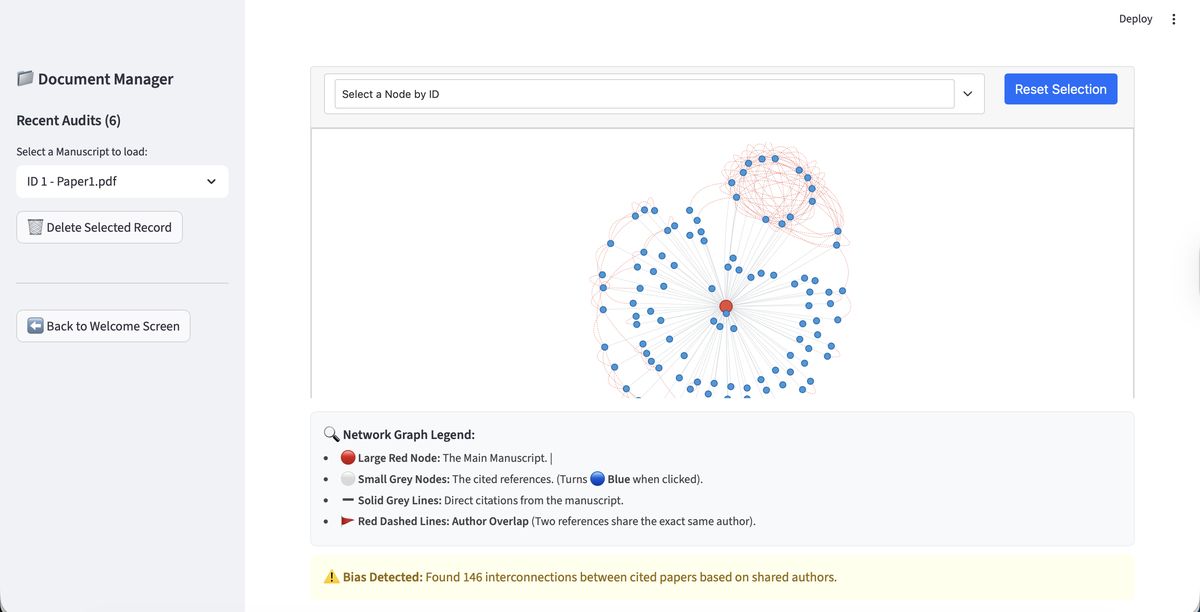}
\caption{Interactive citation-network visualization (manuscript--reference links and shared-author interconnections).}
\label{fig:a15}
\end{figure}

\section{Extended Technology Stack}
\label{app:tech}

\begin{table}[H]
\centering
\caption{Core technologies used in the CitePrism prototype.}
\label{tab:a1}
\begin{tabular}{ll}
\toprule
\textbf{Component} & \textbf{Technology} \\
\midrule
PDF parsing LLM & Google Gemini 2.5 Flash \\
Scoring LLM & \texttt{Llama-3.1-8B-Instruct} (HuggingFace) \\
Sentence embeddings & \texttt{all-MiniLM-L6-v2} \\
Metadata primary source & OpenAlex API \\
Metadata fallback & Semantic Scholar, Crossref, arXiv, web retrieval \\
Fuzzy matching & \texttt{thefuzz} (Levenshtein ratio) \\
Network visualization & PyVis, NetworkX \\
State management & SQLite (\texttt{database/citeprism.db}) \\
Report generation & \texttt{fpdf2} (PDF), HTML \\
Frontend & Streamlit 1.53.0, Plotly \\
\bottomrule
\end{tabular}
\end{table}

\section{Threshold Guidance}
\label{app:threshold}

\textbf{Three-band relevance} is defined on $RS_{\mathrm{final}}$: Relevant ($\geq 70$), Borderline ($40 \leq RS_{\mathrm{final}} < 70$), Irrelevant ($< 40$). These bands support interpretive reading of scores.

\textbf{Operating threshold $\tau$} controls binary Flagged vs.\ Clean triage in the analyst interface and evaluation module. Lower $\tau$ increases sensitivity (more flags); higher $\tau$ reduces analyst workload. In the Paper~1 pilot, $\tau = 17$ yielded $\kappa = 0.429$ with full recall on human-labeled irrelevant citations in that run. Journals should calibrate $\tau$ against editorial policy (e.g., integrity screening vs.\ routine triage).

\begin{table}[H]
\centering
\caption{Threshold sensitivity in the Paper~1 pilot ($n=104$). Row $\tau=17^{\mathrm{a}}$ matches the primary results in Table~\ref{tab:results}. Other rows are recomputed from the public repository scored export (\texttt{Paper1\_scored.json}) and \texttt{gold\_labels\_paper1.csv} to illustrate sensitivity; they are not independent validation runs.}
\label{tab:threshold_sensitivity}
\small
\begin{tabular}{rrrrrrrrr}
\toprule
$\tau$ & Acc. & P(Flag) & R(Flag) & F1(Flag) & Macro-F1 & Wtd-F1 & $\kappa$ & \#Flagged \\
\midrule
10 & 0.808 & 0.875 & 0.269 & 0.412 & 0.648 & 0.767 & 0.333 & 8 \\
15 & 0.837 & 0.765 & 0.500 & 0.605 & 0.751 & 0.824 & 0.507 & 17 \\
17$^{\mathrm{a}}$ & 0.721 & 0.420 & 1.000 & 0.592 & 0.690 & 0.749 & 0.429 & 50 \\
20 & 0.808 & 0.615 & 0.615 & 0.615 & 0.744 & 0.808 & 0.487 & 26 \\
25 & 0.817 & 0.606 & 0.769 & 0.678 & 0.775 & 0.824 & 0.553 & 33 \\
\bottomrule
\end{tabular}
\end{table}

\section{Hybrid Relevance Scoring Pseudo-code}
\label{app:pseudocode}

\begin{lstlisting}
for each reference r in manuscript.references:
    context = extract_citation_context(r)
    meta = enrich_metadata(r)  # OpenAlex + fallbacks
    RS_embed = cosine(embed(manuscript.abstract), embed(meta.abstract))
    RS_llm = llm_score(manuscript.abstract, context, meta.abstract)
    RS_final = 0.6 * RS_llm + 0.4 * RS_embed
    band = categorize(RS_final)  # Relevant / Borderline / Irrelevant
    flags = integrity_checks(r, meta, RS_final)
    binary_flag = (RS_final < tau)
\end{lstlisting}

\section{Sample Audit Report Schema}
\label{app:schema}

\begin{lstlisting}
{
  "manuscript_id": "hash",
  "reference_id": "ref_042",
  "RS_final": 28.5,
  "RS_llm": 22.0,
  "RS_embed": 38.2,
  "band": "Irrelevant",
  "flagged_at_tau": true,
  "tau": 17,
  "intent": "background",
  "rationale": "...",
  "flags": ["MISSING_ABSTRACT"],
  "self_cite": false
}
\end{lstlisting}

\end{document}